\documentclass[aps,prl,preprint,superscriptaddress,floatfix]{revtex4-2}
\usepackage{amsmath,amssymb,graphicx,bm}
\usepackage[colorlinks=true,linkcolor=blue,citecolor=blue,urlcolor=blue]{hyperref}
\graphicspath{{figs/}}

\newcommand{\ExpVEmax}{0.057}
\newcommand{\ExpViolating}{0.422}

\newcommand{\ExpDMcontrol}{0.011}

\newcommand{\ExpDMthixo}{0.165}
\newcommand{\ExpDMfold}{15}
\newcommand{\ExpDMresidHi}{28}
\newcommand{\FloorTwo}{0.063}

\newcommand{\Threshold}{0.13}

\newcommand{\FloorTauThree}{0.056}

\newcommand{\FloorTauTen}{0.106}
\newcommand{\FloorTauThirty}{0.171}

\newcommand{\AmpMu}{1}
\newcommand{\AmpLomaxwell}{0.063}

\newcommand{\AmpLosaramito}{0.017}
\newcommand{\AmpHisaramito}{0.779}

\newcommand{\AmpLothixove}{0.166}

\newcommand{\AmpLomoore}{0.165}

\newcommand{\AmpLolwfinite}{0.184}

\newcommand{\AmpLoaging}{0.106}

\newcommand{\MuPeakthixove}{0.165}
\newcommand{\MuPeakLocthixove}{1.06}

\newcommand{\MuPeakLocmoore}{0.79}

\newcommand{\MuPeakLoclwfinite}{2.70}

\newcommand{\Tstarthixove}{2137}

\newcommand{\Tstarmoore}{1589}

\newcommand{\RidgeConst}{2.9}
\newcommand{\PhasemapMax}{0.33}
\newcommand{\PhasemapMyMinDetect}{5.2}
\newcommand{\ThermoStart}{0.063}
\newcommand{\ThermoEnd}{0.0064}

\newcommand{\GiesWiTen}{0.072}
\newcommand{\FenepWiTen}{0.079}

\newcommand{\CtrlfenepExcessMax}{0.102}

\newcommand{\CtrlgiesekusExcessMax}{0.032}

\newcommand{\CtrlthixoveExcessMax}{0.254}

\newcommand{\ConceptMaxwell}{0.063}
\newcommand{\ConceptThixo}{0.165}
\newcommand{\ConceptThixoMu}{0.92}

\newcommand{\AgingExcessMidEnd}{0.433}

\newcommand{\dkk}{\Delta_{\mathrm{KK}}}      
\newcommand{\dfit}{\Delta_{\mathrm{fit}}}    
\newcommand{\Mu}{\mathrm{Mu}}                
\newcommand{\My}{\mathrm{My}}                
\newcommand{\Wi}{\mathrm{Wi}}                
\newcommand{\De}{\mathrm{De}}                
\newcommand{\gd}{\dot\gamma}
\newcommand{\Gs}{G^{*}}

\begin{document}

\title{Thixotropy versus viscoelasticity: a matter of time}
\author{Rishabh V. More}
\email{rishabh.more@monash.edu}
\affiliation{Bioresource Processing Institute of Australia (BioPRIA), Department of Chemical and Biological Engineering, Monash University, Clayton, VIC 3800, Australia}
\date{\today}

\begin{abstract}
Thixotropic and viscoelastic materials both remember their deformation history and are hard to tell
apart. However, a stationary linear response obeys the Kramers--Kronig relations. A structure that
evolves during a sweep leaves the spectrum inconsistent with them, so their residual, from one chirp and
no model, measures nonstationarity. It survives the small-amplitude limit, returns to its
baseline at any stationary state, and for bounded restructuring peaks near the structural time. We verify
this on eight models and demonstrate it on published spectra of a thixotropic silica suspension.
\end{abstract}
\maketitle

Many soft materials such as colloidal gels, clays, blood, and drilling fluids remember how they were
deformed \cite{larsonwei,mewiswagner,coussot}. This memory can have two different
physical origins, and the two have long been difficult to tell apart \cite{larsonwei,barnes}. In a
\emph{viscoelastic} liquid the memory is elastic, i.e., polymer chains or transient networks store
deformation energy and release it over a relaxation time $\tau_{\rm ve}$. In a \emph{thixotropic}
material the memory is structural, i.e., flow breaks down a microstructure that rebuilds at rest over a
thixotropic time $\tau_{\rm thix}$, so that the viscosity itself changes with time
\cite{larsonwei,mewiswagner}. Both shear thin and both show history dependent transients \cite{larsonwei,agarwal}, so the two are
easily confused.

The two are conventionally described by different dimensionless groups. Viscoelasticity is governed by
the Weissenberg number $\Wi=\gd\,\tau_{\rm ve}$, with $\gd$ the imposed shear rate, and by the Deborah
number $\De=\tau_{\rm ve}/t_{\rm obs}$, with $t_{\rm obs}$ the observation time \cite{reiner}. Thixotropy
needs two more, the Mnemosyne number $\My=\gd\,\tau_{\rm thix}$ and a mutation number
$\Mu=t_{\rm obs}/\tau_{\rm thix}$, which Jamali and McKinley combined into a two-dimensional map of
thixotropic behavior \cite{mnemosyne,mours}. A material is nonlinearly viscoelastic when $\Wi\gtrsim1$
and thixotropic when its structure evolves, i.e., when $\My$ and $\Mu$ are of order one. The existing
methods to distinguish the two are transient tests such as step-down and start-up flows, step strain or
step stress after preshear \cite{agarwal}, and hysteresis loops in shear-rate ramps
\cite{wangewoldt,divoux}, several of which have proved ambiguous \cite{agarwal}.

In this paper, we show that the difference between the two kinds of memory is a symmetry of the memory
kernel that can be tested with small-amplitude oscillatory measurements. We write the linearized constitutive law as a kernel $G$ that relates the shear stress $\sigma$ to the
history of the strain rate,
\begin{equation}
\sigma(t)=\int_{-\infty}^{t}\! G\!\left(t-t',\,t'\right)\,\gd(t')\,dt' ,
\label{eq:kernel}
\end{equation}
where $t$ is the present time and $t'$ is the time of an earlier deformation. For a viscoelastic material
at rest or at a steady state the kernel depends only on the elapsed time $t-t'$, so that the properties
of the material are the same no matter when the clock is started. We call this case
time-translation invariance (TTI). For a thixotropic or aging material the kernel depends on both times,
because the structure changes during the time in which the response is recorded \cite{fsc}. TTI holds
once the response has settled to a stationary state, i.e., a state whose properties do not change in
time, and it is broken while the material evolves during the measurement, whether the structure is
building up or breaking down, or whether a nonlinear viscoelastic conformation is still relaxing from an
earlier deformation.

TTI has a signature in the frequency domain. For a causal and time-translation-invariant linear response,
the storage modulus $G'(\omega)$ and the loss modulus $G''(\omega)$, i.e., the real and imaginary parts
of the complex modulus $\Gs(\omega)$ at angular frequency $\omega$, are related to each other by the
Kramers--Kronig (KK) integral relations \cite{booij,kwon,shanbhag}. When the material evolves during a
frequency sweep, the measured $G'$ and $G''$ come from a material that was different at different
frequencies, and the resulting finite-window spectrum will generally not be reproducible by a single
stationary modulus. Causality alone keeps a
generalized form of the relations exact at each instant \cite{solis}, so what fails is the description of
the data collected over the finite measurement window by one stationary spectrum. Fielding, Sollich and
Cates noted that under aging these relations should be verified and not assumed \cite{fsc}, and
Ramya \emph{et al.} \cite{ramya} showed that a KK residual detects the loss of time independence in a
thixotropic suspension. However, a single residual cannot separate nonlinearity from time dependence.
The aim of this work is to turn this residual into a quantitative measure of nonstationarity. To this end, we calibrate its floor, and we then show that an
amplitude sweep removes nonlinear distortion, that the residual returns to its floor at any stationary
state, and that a sweep of the chirp duration locates the time scale of the structural change. A single residual only reports that the response changed within the measurement window, and
distinguishing thixotropy from viscoelasticity is the result of the full protocol, i.e., the residual
together with its dependence on amplitude, duration, and flow history.

\begin{figure}[t]
\includegraphics[width=\columnwidth]{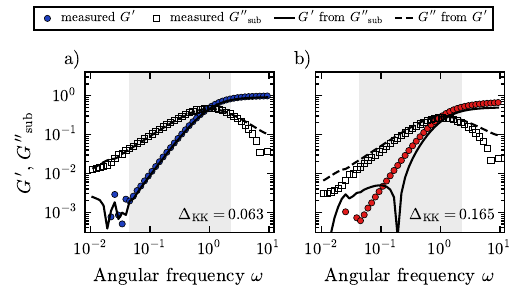}
\caption{Moduli measured from one chirp (symbols) and their Kramers--Kronig reconstructions from
Eqs.~(\ref{eq:kk}) and (\ref{eq:kk2}) (lines). The shading marks the central $60\%$ of the measured
frequency range that enters Eq.~(\ref{eq:delta}). (a)~A Maxwell fluid is time-translation invariant, so
the reconstructions and the measurements agree and $\dkk=\ConceptMaxwell$ is the floor. (b)~A
thixo-viscoelastic fluid whose structure rebuilds during the sweep ($\lambda_0=0.2$, $\beta=1$,
$\Mu=\ConceptThixoMu$). The storage modulus reconstructed from the loss modulus misses the measured one
and $\dkk=\ConceptThixo$. In both cases $\gamma_0=10^{-6}$.}
\label{fig:concept}
\end{figure}

\emph{The residual.} From one measured spectrum we reconstruct each modulus from the other using the
subtracted KK transforms \cite{booij,ramya},
\begin{align}
G''_{\rm rec}(\omega)&=\frac{2\omega}{\pi}\,\mathrm{P}\!\!\int_0^\infty\!
\frac{G'(u)-G'(\omega)}{u^2-\omega^2}\,du ,
\label{eq:kk}\\
G'_{\rm rec}(\omega)&=-\frac{2\omega^2}{\pi}\,\mathrm{P}\!\!\int_0^\infty\!
\frac{G''_{\rm sub}(u)/u-G''_{\rm sub}(\omega)/\omega}{u^2-\omega^2}\,du ,
\label{eq:kk2}
\end{align}
where $u$ is the integration frequency, $\mathrm{P}$ denotes the Cauchy principal value, and the
subscript ``rec'' marks a reconstructed modulus. $G''_{\rm sub}=G''-\eta_\infty\omega$ is the measured
loss modulus with the Newtonian solvent contribution removed, where
$\eta_\infty=\lim_{\omega\to\infty}G''/\omega$ is the high-frequency viscosity. The subtractions in
Eqs.~(\ref{eq:kk}) and (\ref{eq:kk2}) remove the integrable singularity as well as any constant
equilibrium modulus $G'(0)=G_e$, so that a yield-stress solid needs no special treatment. We compare each
reconstruction with its measured counterpart over the central $60\%$ of the measured frequency range,
where the error from truncating the integrals to the measured range is small, and combine the two
normalized mismatches as
\begin{equation}
\dkk^2=\frac{\lVert G''_{\rm rec}-G''_{\rm sub}\rVert^2+\lVert G'_{\rm rec}-(G'-G_e)\rVert^2}
{\lVert (G',G'')\rVert^2},
\label{eq:delta}
\end{equation}
with $G_e$ taken as the storage modulus at the lowest measured frequencies. In this way, both directions
of the KK pair are tested by one dimensionless number. Figure~\ref{fig:concept} shows the
reconstructions for a Maxwell fluid and for a thixo-viscoelastic fluid whose modulus rebuilt during the
sweep. An exactly KK-consistent spectrum does not give zero but a small value that we call the floor,
$\dkk^0$. The floor is the residual of a spectrum that satisfies the relations exactly, and it is set by
the discretization and by the finite number of frequency decades measured and not by the material
\cite{SM}. Through the full chirp pipeline described next the floor is $\dkk^0=\FloorTwo$, and we use
$2\dkk^0=\Threshold$ as the detection threshold throughout. This threshold is a working choice and not a
statistical confidence limit, and it has to be recalibrated for the frequency range, the sampling, the
noise level, and the implementation of the residual used in a given experiment. The floor also depends on
where a relaxation mode lies with respect to the measured frequency range, and for a single stationary
mode it varies from $\FloorTauThree$ to $\FloorTauThirty$ as the relaxation time moves from the center
of the measured range to beyond its low-frequency end \cite{SM}. Hence, wherever the relaxation time of a
model leaves the measured range, we quote the floor of a stationary mode with the same relaxation time,
which we call the matched floor.

\emph{Chirp measurement.} A small residual is meaningful only if the whole spectrum is captured before
the material evolves, which is achieved by optimally windowed chirps (OWCh) \cite{geri}. A single
exponential chirp has the strain $\gamma(t)=\gamma_0\sin\phi(t)$ with amplitude $\gamma_0$, and its
instantaneous frequency sweeps from $\omega_1$ to $\omega_2$ over a duration $T$. One sweep returns
$\Gs(\omega)$ over the whole frequency range in a time short compared with $\tau_{\rm thix}$, i.e., at a
mutation number $\Mu=T/\tau_{\rm thix}$ that can be pushed below one \cite{mnemosyne,geri}. We use a
three-decade sweep from $\omega_1=10^{-2}$ to $\omega_2=10$ with $T=4000$ unless stated otherwise
\cite{SM}. Everything downstream of the stress signal uses no constitutive information. We validate the test on eight standard models, which we integrate as conformation
tensors with a fourth-order Runge--Kutta scheme \cite{SM}. These are the linear Maxwell/Oldroyd-B,
FENE-P, and Giesekus models (viscoelastic); the Saramito elasto-visco-plastic (EVP) model with yield
stress $\sigma_y$ \cite{saramito}; the inelastic Moore model with viscosity
$\eta(\lambda)=\eta_s(1+9\lambda)$ set by a structure parameter $\lambda\in[0,1]$ \cite{larsonwei}; a
thixo-viscoelastic Maxwell mode with modulus $G_0\lambda(t)$; a bounded (Larson--Wei) aging model whose
relaxation time saturates at $10\tau_{\rm ve}$; and an unbounded aging Maxwell mode \cite{larsonwei}.
Units are $\tau_{\rm ve}=G_0=1$ with $\eta_s=0.1$ \cite{SM}. The structure parameter obeys $\dot\lambda=(1-\lambda)/\tau_{\rm thix}-\beta\lambda|\gd|$
with breakdown coefficient $\beta$. Because the breakdown term is linear in $|\gd|$, the rectified rate
of the chirp itself lowers the mean structure at first order in $\gamma_0$, and not at second order as a
smooth nonlinearity would. Hence, the probe leaves the structure alone only when
$\gamma_0\,\omega_2\,\tau_{\rm thix}\ll1$. We use $\gamma_0=10^{-6}$ for every small-amplitude sweep,
which keeps this product below $0.6$ over the whole range of $\tau_{\rm thix}$ studied and below $0.04$
for $\Mu\ge1$. For $\Mu<1$ the structure lost to the probe over the whole sweep stays under half a percent \cite{SM}. In an experiment, the small-amplitude plateau of the residual is the test that this condition has been met.

\begin{figure}[tbp]
\includegraphics[width=\columnwidth]{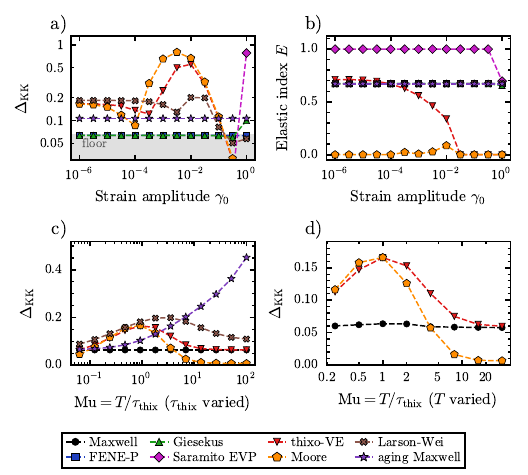}
\caption{The control axes. (a)~$\dkk$ as a function of the strain amplitude $\gamma_0$ at $\Mu=\AmpMu$.
The viscoelastic models stay at the floor (grey) for every $\gamma_0\le0.3$, whereas the restructuring
and aging models level off at a finite value as $\gamma_0\to0$. (b)~Elastic index $E$. (c)~$\dkk$ as a
function of the mutation number $\Mu=T/\tau_{\rm thix}$ at fixed $T$ and $\gamma_0=10^{-6}$. Thixo-VE
and Moore peak near $\Mu\approx1$ and return to the floor, the bounded Larson--Wei model returns to the
floor of its own equilibrium spectrum ($\FloorTauTen$ for a mode at $10\tau_{\rm ve}$), and the
unbounded aging mode rises without returning. (d)~Duration sweep on one material, i.e., $\dkk$ as a
function of $T$ at fixed $\tau_{\rm thix}=2000$; Maxwell gives the floor at each $T$. Parameters: FENE-P
$L^2=50$, Giesekus $\alpha=0.3$, Saramito $\sigma_y=0.5$; restructuring models $\lambda_0=0.2$,
$\beta=1$.}
\label{fig:axes}
\end{figure}

\emph{Amplitude axis.} A nonlinear but structurally stationary material is time-translation invariant.
Hence, its first-harmonic spectrum must return to KK consistency as $\gamma_0\to0$, with the leading
corrections to $G'$ and $G''$ scaling as $\gamma_0^2$. A material that is rebuilding from an
out-of-equilibrium state does so whether or not it is probed, so its residual survives the
small-amplitude limit. Figure~\ref{fig:axes}(a) shows both behaviors. Maxwell, FENE-P and Giesekus stay
at the floor ($\dkk=\AmpLomaxwell$) for all $\gamma_0\le0.3$, and the Saramito EVP model below yield
gives $\AmpLosaramito$. Nonlinearity shows up only above yield ($\gamma_0=1$, $\dkk=\AmpHisaramito$).
In contrast, the thixo-VE, Moore and Larson--Wei models level off at $\dkk=\AmpLothixove$,
$\AmpLomoore$ and $\AmpLolwfinite$ as $\gamma_0\to0$, above the threshold of $\Threshold$, and the aging
mode levels off at $\AmpLoaging$. Their curves are not monotonic. Between
$\gamma_0\approx3\times10^{-4}$ and $10^{-1}$ the probe breaks the structure down and the residual rises
several fold, and for $\gamma_0\ge0.3$ the probe holds the structure broken down for the whole sweep, so
that the response is dominated by the solvent and $\dkk$ drops below the floor. Thus, extrapolating to
$\gamma_0\to0$ separates nonlinearity ($\Wi$) from time dependence ($\Mu$), provided the extrapolation
reaches the plateau.

\emph{Elastic content.} Inelastic (``ideal'') thixotropy and thixo-viscoelasticity both break TTI, but the former has essentially no storage modulus, so we separate them with the index $E=\max G'/(\max G'+\max G''_{\rm sub})\in[0,1]$ over the measured range, which is $\approx0$ for the Moore model, $0.6$--$0.7$ for the viscoelastic and thixo-viscoelastic liquids, and $\to1$ for the Saramito solid below yield, as shown in Fig.~\ref{fig:axes}(b).

\emph{Mutation axis.} The way in which TTI is broken shows in the dependence of $\dkk$ on $\Mu$
[Fig.~\ref{fig:axes}(c)]. Over one oscillation the kernel drifts by $(\partial_tG)(2\pi/\omega)$, so the
part of the spectrum that violates the KK relations is first order in the fractional drift per cycle,
$\omega^{-1}|\partial_t\ln\Gs|$, which is the per-cycle mutation number of Mours and Winter \cite{mours}.
As a result, $\dkk$ grows while the structure evolves and falls back once the structure equilibrates
within the measurement window. Thixo-VE and Moore peak at $\dkk\approx\MuPeakthixove$ near
$\Mu^\star=\MuPeakLocthixove$ and $\MuPeakLocmoore$, respectively, and are at or below the floor by
$\Mu\approx30$. The bounded Larson--Wei model peaks at $\Mu^\star=\MuPeakLoclwfinite$ and settles at the
floor of a stationary mode at its equilibrium relaxation time of $10\tau_{\rm ve}$. The unbounded aging
mode does not turn over, and its residual at $\Mu=100$ is $\AgingExcessMidEnd$ above the matched floor
of a stationary mode with its mid-sweep relaxation time \cite{SM}. From rest only the build-up term of
the kinetics acts, so replacing the rate-controlled breakdown by a stress-controlled one \cite{souza}
leaves the curve unchanged \cite{SM}.

The position of the peak is a practical measure of the restructuring time scale, and a duration sweep
on one material reads it directly. In Fig.~\ref{fig:axes}(d) we hold $\tau_{\rm thix}=2000$ and vary $T$
from $500$ to $64000$, recalibrating the floor with a Maxwell fluid at each $T$. The residual peaks at
$T^\star=\Tstarthixove$ for thixo-VE and $\Tstarmoore$ for Moore, and dividing by the $\Mu^\star$ of
Fig.~\ref{fig:axes}(c) returns $\tau_{\rm thix}$ to within $1\%$ of the value used. $\Mu^\star$ is of
order one here, but it depends on the kinetics and the protocol, so a calibration against a fluid of
known $\tau_{\rm thix}$ is needed before $T^\star$ is read as $\tau_{\rm thix}$ itself. For the aging mode measured at successive ages, the excess over the matched
floor follows the mutation within the measurement window, $T/(t_{\rm age}+t_w)$, and falls with the age
$t_w$ \cite{SM}. Thus, a finite measurement locates the current rate of change but cannot confirm that
the evolution is unbounded.

\begin{figure}[tbp]
\includegraphics[width=\columnwidth]{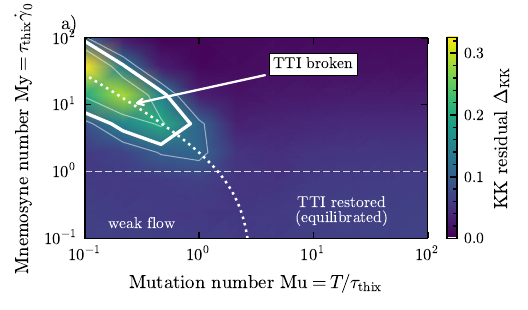}\\[2pt]
\includegraphics[width=\columnwidth]{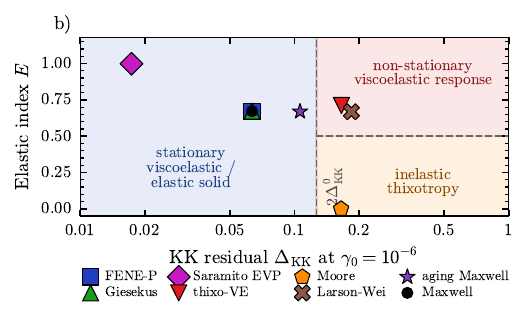}
\caption{(a)~State diagram: $\dkk$ (color) on the $(\Mu,\My)$ plane for a thixo-viscoelastic fluid at
rest equilibrium ($\lambda_0=1$) driven by a steady base shear and measured by an orthogonal chirp of
amplitude $2\times10^{-6}$. White lines: solid, the detection threshold $2\dkk^0=\Threshold$ (faint,
$1.5\dkk^0$ and $3\dkk^0$); dotted, the crest $\Mu(1+\My)\approx\RidgeConst$; dashed, $\My=1$. The
residual reaches $\PhasemapMax$ and clears the threshold for $\My\gtrsim\PhasemapMyMinDetect$ \cite{SM}.
(b)~The models on the ($\dkk$, $E$) plane using the $\gamma_0=10^{-6}$ values of Fig.~\ref{fig:axes}.
Shading and dashed lines mark the threshold and $E=0.5$.}
\label{fig:phase}
\end{figure}

\emph{A state diagram.} Figure~\ref{fig:phase}(a) shows $\dkk$ on the $(\Mu,\My)$ plane for a
thixo-viscoelastic fluid driven from rest equilibrium by a steady shear of strength $\My$ while a chirp
in the orthogonal direction measures its spectrum. Broken TTI occupies a bounded region, since $\dkk$ is at the floor when the flow is too weak to restructure the material ($\My\lesssim1$ \cite{coussot}) and when the structure equilibrates before the sweep ends (large $\Mu$). Flow shortens the structural time toward $\tau_{\rm thix}/(1+\My)$, so the crest of the region follows $\Mu(1+\My)\approx\RidgeConst$ (dotted line). With the
$\gamma_0\to0$ values of Fig.~\ref{fig:axes}, the models fall into four regions of the ($\dkk$, $E$)
plane without any constitutive fitting, as shown in Fig.~\ref{fig:phase}(b). The viscoelastic liquids
and the EVP solid are at the floor, ideal thixotropy is above threshold with $E\to0$, and
thixo-viscoelasticity and bounded aging are above threshold with intermediate $E$, in the region we label a nonstationary viscoelastic response, because aging and, in principle, a relaxing viscoelastic conformation can occupy it as well. The unbounded aging mode at $\Mu=1$ is just below threshold and crosses it as $\Mu$ grows. The axes therefore classify the models
examined here, and in practice they distinguish stationary viscoelasticity, nonlinear viscoelasticity,
inelastic thixotropy, and a restructuring viscoelastic material once the flow-history controls described
below have excluded a viscoelastic conformation that is still relaxing from an earlier deformation.

\begin{figure}[tbp]
\includegraphics[width=\columnwidth]{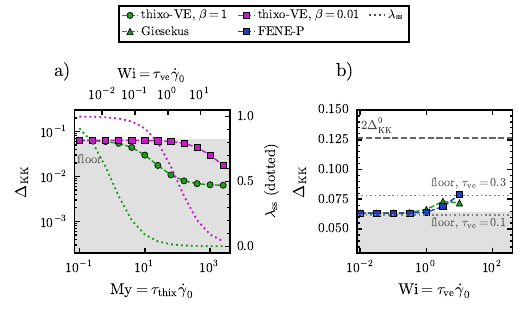}
\caption{Driven steady states. (a)~Thixo-VE fluid under a steady base shear of strength
$\My=\tau_{\rm thix}\gd_0$ (top axis: $\Wi=\tau_{\rm ve}\gd_0$), measured by an orthogonal chirp
($\gamma_0=2\times10^{-3}$). Circles, $\beta=1$; squares, $\beta=0.01$; dotted lines, the steady
structure $\lambda_{\rm ss}$ (right axis). Grey: the floor. (b)~Giesekus and FENE-P under a steady base
flow of Weissenberg number $\Wi$. Dashed line, the threshold; dotted lines, the floor of a stationary
mode at $\tau_{\rm ve}=0.3$ and $0.1$. Parameters: $\tau_{\rm thix}=50$, $T=4000$.}
\label{fig:thermo}
\end{figure}

\emph{A stationarity result.} The per-cycle drift of the kernel ties $\dkk$ to the rate at which the
spectrum changes during the measurement, and not to the steady dissipation. A material under a constant
drive dissipates steadily, yet once its structure reaches a steady state ($\langle\dot\lambda\rangle=0$)
the spectrum stops changing. Hence, $\dkk$ should fall to the floor at \emph{any} stationary state, in
equilibrium or under a steady drive. We test this in Fig.~\ref{fig:thermo} with an orthogonal chirp
superposed on a steady base shear \cite{SM}. For the
thixo-VE fluid with $\beta=1$ the residual is at the floor at weak flow ($\ThermoStart$ at $\My=0.1$)
and falls below it as $\My$ grows, reaching $\ThermoEnd$ at $\Wi=50$. The fall follows the steady
structure $\lambda_{\rm ss}=1/(1+\beta\My)$ and not the flow strength, since with $\beta=0.01$ the
residual stays at the floor up to $\Wi\approx1$, where $\lambda_{\rm ss}\ge0.6$. Below the floor, the
structural modulus is small compared with the solvent contribution and the normalization in
Eq.~(\ref{eq:delta}) shrinks the residual, and at no steady state does it rise. The nonlinear viscoelastic
models under a steady base flow [Fig.~\ref{fig:thermo}(b)] give $\GiesWiTen$ (Giesekus) and
$\FenepWiTen$ (FENE-P) at $\Wi=10$, which is below the threshold and within the range of floors of a
stationary mode whose relaxation time has been shortened by shear thinning into the upper part of the
measured range \cite{SM}. This is the reason why the mutation axis returns to the floor once the
structure equilibrates and why the detectable region in Fig.~\ref{fig:phase}(a) is bounded.

The residual does not tell us the origin of the evolution it reports, so the flow-history controls are
important. A presheared FENE-P or Giesekus fluid probed while its conformation relaxes after the flow is
stopped gives, once the matched floor is removed, an excess of at most $\CtrlfenepExcessMax$ (FENE-P)
and $\CtrlgiesekusExcessMax$ (Giesekus), which is below the threshold, compared with
$\CtrlthixoveExcessMax$ for the thixo-VE fluid relaxing on its structural time scale in the same protocol
\cite{SM}. We do not take this result as general, so a residual above threshold is assigned to structure only
after the equilibrium and steady-drive controls above have been run.

\begin{figure}[tbp]
\includegraphics[width=\columnwidth]{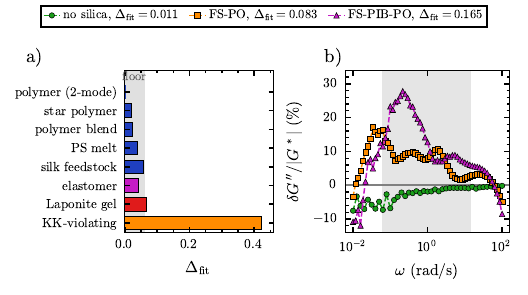}
\caption{Experimental data. (a)~Published spectra grouped by class. Viscoelastic polymers and an
elastomer lie in the floor region (grey, $\dfit\le\ExpVEmax$), whereas a control spectrum that we made
KK-inconsistent on purpose is flagged (orange). Sources: pyReSpect dataset \cite{poudel} and RepTate \cite{reptate}; PS, polystyrene. (b)~The Dullaert--Mewis model thixotropic fluid (fumed silica in PIB/paraffin oil), from
the small-amplitude frequency sweeps of Ramya \emph{et al.} \cite{ramya} over four decades, showing the
frequency-resolved residual of the loss modulus $\delta G''(\omega)/|\Gs|$. The silica-free matrix
(green) is at the floor, whereas adding fumed silica makes the response thixotropic and the residual
systematic (FS-PO, orange; full FS-PIB-PO suspension, purple). Strains of $0.25$--$1\%$ (linear regime).
Grey shading marks the central $60\%$ of the frequencies used for $\dfit$. Method: one non-negative
Maxwell fit with about $3$ modes per decade \cite{SM}.}
\label{fig:exp}
\end{figure}

\emph{Demonstration on experimental data.} To show the test on real measurements, we apply it to
published spectra (Fig.~\ref{fig:exp}). The transforms in Eqs.~(\ref{eq:kk}) and (\ref{eq:kk2}) amplify
noise, and the reconstruction of $G'$ from $G''$ is ill-conditioned when $G''\ll G'$, as in a stiff gel.
For measured spectra we therefore use a related residual. We fit one Maxwell spectrum with all weights
held non-negative to $G'$ and $G''$ at the same time, and take the mismatch of this joint fit, normalized
as in Eq.~(\ref{eq:delta}), as the residual $\dfit$. A single such spectrum can match both parts only
when they obey the KK relations, so a mismatch at the floor implies KK consistency, whereas the converse
is stricter because non-negative weights also impose passivity and a finite number of modes
\cite{ramya,poudel}. On the simulated spectra at $\Mu=1$ the two residuals give the same classification for the
restructuring models, but $\dfit$ is less sensitive, in particular to a drift of the relaxation time,
and does not reproduce every $\dkk$ classification, so we keep separate symbols and compare the two in
\cite{SM}. The method
leaves ordinary viscoelastic materials (a polymer blend, a star polymer, and a polystyrene melt
\cite{poudel,reptate}) and an elastomer with a terminal plateau within the floor region
($\dfit\le\ExpVEmax$), and it detects a control spectrum that we deliberately made KK-inconsistent by
adding a mismatched extra mode to $G''$ alone, at $\dfit=\ExpViolating$ [Fig.~\ref{fig:exp}(a)]. For an
experimental thixotropic fluid we use the Dullaert--Mewis system of fumed silica in a polyisobutylene/paraffin oil matrix \cite{dullaert}, measured by Ramya \emph{et al.} \cite{ramya} [Fig.~\ref{fig:exp}(b)]. The silica-free matrix is KK-consistent with
$\dfit=\ExpDMcontrol$, i.e., at the floor. Adding $2.9$~vol\% of fumed silica makes the suspension
thixotropic and the residual \emph{systematic}, i.e., a coherent deviation of one sign reaching
${\approx}\ExpDMresidHi\%$ of $|\Gs|$, with $\dfit=\ExpDMthixo$, some $\ExpDMfold$ times the matrix
floor. The sweeps were taken at small strains ($0.25$--$1\%$, in the linear regime), so nonlinearity is
excluded. Ramya \emph{et al.} attributed the violation to the time dependence of the thixotropic
structure, i.e., to broken TTI, and we read it as a quantitative measure of nonstationarity \cite{SM}.
We note that the measure is one-sided. A residual above the floor indicates a spectrum that evolved during the
sweep, once the amplitude axis has excluded nonlinear distortion, whereas a residual at the floor is
necessary but not sufficient for stationarity, since an evolution that kept $G'$ and $G''$ mutually
consistent would not be detected.

In summary, we have established the residual on eight standard constitutive models and demonstrated it
on published spectra. The protocol, i.e., the amplitude, chirp-duration, and flow-history sweeps, needs
only small-amplitude chirps on a commercial rheometer with the OWCh input \cite{geri}. We expect it to be most useful for weakly elastic
pastes and gels, where transient tests have proved ambiguous \cite{agarwal}, and the natural next step
is to run the duration sweep of Fig.~\ref{fig:axes}(d) on one restructuring fluid. Since the residual reports any evolution within the measurement window, assigning it to structure
rests, as here, on the flow-history controls.

\begin{acknowledgments}
\end{acknowledgments}

\end{document}